\documentstyle[12pt]{article}
\begin{document}
 \title{Stronger-than-quantum correlations}
\author{G. Krenn\\
 {\small Atominstitut der \"Osterreichischen Universit\"aten}  \\
  {\small Sch\"uttelstra\ss e 115}    \\
  {\small A-1020 Vienna, Austria   }            \\
{\small and}\\
K. Svozil\\
 {\small Institut f\"ur Theoretische Physik}  \\
  {\small Technische Universit\"at Wien   }     \\
  {\small Wiedner Hauptstra\ss e 8-10/136}    \\
  {\small A-1040 Vienna, Austria   }            \\
  {\small e1360dab@awiuni11.edvz.univie.ac.at}}
\maketitle

\begin{abstract}
After an elementary derivation of Bell's inequality, several forms of
expectation functions for two-valued observables are discussed.
Special emphasis is given to hypothetical stronger-than quantum
expectation functions which give rise to a maximal violation of
Bell's inequality.
\end{abstract}

\noindent
In the following we consider a general two-particle experiment, which is
specified by the measurements performed on the individual particles.
On the first particle a measurement of the dichotomic
(two-valued) observable
$r_\alpha \in \{-1,1\}$ (e.g. the spin direction of a spin-1/2 particle
such as an electron along the unit vector
$\alpha$) is made by observer A,
where $\alpha$ is a measurement parameter
(e.g. a direction); likewise
the dichotomic observable $r_\beta \in \{-1,1\}$
is measured on the second particle by experimenter B.
Then an expectation function can be defined by
\begin{equation}
E(\alpha ,\beta )=\langle r_\alpha r_\beta \rangle =\lim_{N\rightarrow
\infty}{1\over N}
\sum  r_\alpha r_\beta \quad .
\label{e-1}
\end{equation}
Classically,
observer A may perform
measurements of $r_\alpha$ and $r_{\alpha '}$
on particle 1, one after the other; whereas
observer B may perform
successive measurements of $r_\beta$ and $r_{\beta '}$
on particle 2.
(Notice that classically observer A (B)
may perform an arbitrary number of successive measurements on
particle 1 (2) without modifying the state thereof.)
Consider a series of $N$ experiments.
In each experiment the values of
$r_\alpha$ and
$r_{\alpha '}$
($r_\beta$ and
$r_{\beta '}$) are successively measured by observer  A (B) on
particle 1 (2).
\begin{figure}
\begin{center}
Figure 1
\end{center}
\caption{
Four lists of results of
classical experiments performed by observer A
(recording the values of the observables $r_\alpha$ and
$r_\alpha '$ subsequently) and observer B
(recording the values of the observables $r_\beta$ and
$r_\beta '$ subsequently).
``$+$'' stands for $+1$,
``$-$'' stands for $-1$.
 \label{f-1}}
\end{figure}
In such a way one obtains four lists containing the recordings of
the measurement results with
the parameters set to $\alpha$ and $\alpha '$ for particle 1 and
$\beta$ and $\beta '$ for particle 2, as shown in Fig. \ref{f-1}.

Analyzing these results, the two experimenters A and B may now determine
the number of different signs (results) in the four pairs of lists
$(\alpha,\beta )$, $(\alpha ,\beta ')$, $(\alpha ',\beta )$ and
$(\alpha ',\beta ')$.
Let $n(\alpha ,\beta )$ stand for the number of different
signs (results) in the two lists corresponding to the measurements with
the parameters adjusted to $\alpha$ and $\beta$.
Having determined the four values
$n(\alpha ',\beta )$,
$n(\alpha ,\beta )$,
$n(\alpha ,\beta ')$ and
$n(\alpha ',\beta ')$
(cf. Fig. \ref{f-1}), we make a simple observation \cite{krenn}.

Following the ``inner path''
$\alpha '
\rightarrow
\beta
\rightarrow
\alpha
\rightarrow
\beta  '$
from list $\alpha '$ to list $\beta '$ in Fig. \ref{f-1} we have to
change
$n(\alpha ',\beta )$ signs in the first step to get list $\beta$,
$n(\alpha ,\beta )$ signs in the second step to get list $\alpha$,
and $n(\alpha ,\beta ')$ signs in the last step to obtain list $\beta '$.
In such a way the {\em maximum number} of signs we may have changed in
list
$\alpha '$ to obtain list $\beta '$ is simply
$n(\alpha ',\beta )+
n(\alpha ,\beta )+
n(\alpha ,\beta ')$.
But since $n(\alpha ',\beta ')$ is just the number of different signs
in lists $\alpha '$ and $\beta '$
---corresponding to the ``outer path'' in Fig. \ref{f-1}---
$n(\alpha ',\beta ')$ must be smaller than or equal to
the maximum number of sign changes along the inner path
$n(\alpha ',\beta )+
n(\alpha ,\beta )+
n(\alpha ,\beta ')$.
This can be expressed by the inequality
\begin{equation}
n(\alpha ',\beta )+
n(\alpha ,\beta )+
n(\alpha ,\beta ') \ge
n(\alpha ',\beta ')\quad .
\label{e-2}
\end{equation}
The probability for different signs (results) after $N$
experiments can be approximated by
$P^{\neq} (\alpha , \beta )=n(\alpha ,\beta)/N$.
Thus,
the probability for equal signs (results) after $N$
experiments is approximately given by
$P^=(\alpha , \beta )=1-P^{\neq } (\alpha ,\beta)$.
By definition (\ref{e-1}), the expectation value is
\begin{equation}
E(\alpha ,\beta )=
P^=(\alpha , \beta )-P^{\neq } (\alpha ,\beta )=2P^=(\alpha ,\beta
)-1 \quad.
\label{e-1a}
\end{equation}
With these identifications, Eq. (\ref{e-2}) can easily be rewritten into
Bell's inequality
\begin{equation}
-2 \le E(\alpha ',\beta )+
E(\alpha ,\beta )+
E(\alpha ,\beta ') -
E(\alpha ',\beta ')\le 2\quad .
\label{e-3}
\end{equation}
The bound from below can be derived by a similar argument, considering
the number of equal signs (results) $u(\alpha ,\beta )= N-n(\alpha
,\beta )$
instead of the number of different signs (results).   $u(\alpha ,\beta
)$ satisfies the same inequality
(\ref{e-2}) as $n(\alpha ,\beta )$.

So far,
we have been dealing
with arbitrary two-valued
observables.
We shall now concentrate on angular momentum.
Two observers A and B measure the {\em sign} of
the angular momentum of the first and second particle, respectively.
I.e., the observers are only interested in the direction of the angular
momentum
(spin, polarization) and not in its absolute value.
If an observer registers angular momentum ``up,''
the observer assigns the value ``$+1$'' (or ``$+$'') to the event;
``down'' corresponds to ``$-1$'' (or ``$-$'').
We shall thus denote by $r_\alpha \in \{-1,1\}$
the direction of angular momentum measured by observer A
along the unit vector $\alpha$; likewise $r_\beta \in \{-1,1\}$ is
the direction of angular momentum measured by
observer B along the unit vector $\beta$.
Let $\theta =\cos ^{-1} (\alpha \cdot \beta )$ be the angle between
$\alpha$ and
$\beta$.
Assume further that
the particles have been prepared in
a singlet state, i.e.,
that the total angular momentum of the two particles is zero.
For simplicity, it is assumed that
$\alpha$ and $\beta$ lie in the plane perpendicular to
the momentum of the particles.

The assumption of local realism (e.g., \cite{clauser,peres-book})
implies that the
classical expectation function computed from the classical observables
$r_\alpha ={\rm sgn} (\alpha \cdot j^A)$ and
$r_\beta ={\rm sgn} (\beta \cdot j^B)$ (here, $j^{A,B}$ are the classical
angular momenta of the individual particles, as measured by A and B)
is given by
\cite{peres}
\begin{equation}
E(\theta )=2 \theta /\pi -1 \quad .
\label{e-cef}
\end{equation}
This corresponds to the assumption that the average number of
unchanged signs is proportional to the relative angle
$\theta$;
 cf. Eq.
(\ref{e-1a}), i.e.,
\begin{equation}
P^=(\theta )= \theta /\pi \quad .
\end{equation}
 This dependence on the relative angle $\theta$ one would classically
expect from conservation of angular momentum: if $\theta=\pi$, i.e.,
if two observers measure along the same directions,
then there are only two possible outcomes, namely $+\,-$ or $-\,+$
(and $P^=(0)=0$);
if they measure along opposite directions,
then again there are only two possible outcomes, namely $+\,+$ or $-\,-$
(and $P^=(\pi )=1$);
for all the other directions, the outcomes $+\,+$ or $-\,-$
occur with a fraction thereof.

It is one puzzling feature of
quantum mechanics that this no longer holds.
Quantum
mechanically, for two particles
of spin
$j$
in a singlet state,
the
correlation
function is given by (cf. appendix and ref. \cite{gisin-peres})
\begin{equation}
C(\theta )=
- {j(j+1)\over 3} \, \cos \theta \quad .
\end{equation}
To be comparable to the classical expectation function,
the quantum expectation function must be normalized
such that $E_{qm}(\pi )=-E_{qm}(0)=1$;
thus
$E_{qm}(\theta ) = {3/[j(j+1)]} C(\theta )$.
(For electrons, this amounts to a
multiplication of the quantum correlation function $C$
by a factor of 4.)
Thus, for an arbitrary particle with nonvanishing spin, the quantum
mechanical expectation function is given by
\begin{equation}
E_{qm}(\theta )=-\alpha \cdot \beta =-\cos \theta \quad .
\label{e-qm}
\end{equation}

As a result,
for angles $\pi /2< \theta <\pi$,
the pair of outcomes $+\,+$ or $-\,-$ occurs more often,
the pair of outcomes $+\,-$ or $-\,+$ occurs less often;
for angles $0< \theta <\pi /2$,
the pair of outcomes $+\,+$ or $-\,-$ occurs less often,
the pair of outcomes $+\,-$ or $-\,+$ occurs more often
than predicted by classical
theories under the assumption of local realism.
To put it pointedly:
for almost all angles $\theta$,
the quantum mechanical expectations are {\em stronger} than the
classical ones.


Note that is is always possible to rewrite any one-to-one expectation
function
$E_\ast (\theta )$ in terms of the classical expectation
function and {\it vice versa}.
E.g.,
for $0\le \theta \le \pi$,
insertion of Eq.
(\ref{e-cef}) into Eq.
(\ref{e-qm}) yields
\begin{equation}
E_{qm}(\theta )=-\cos ({\pi \over 2}(E(\theta )+1))
,\quad
E(\theta )=2\cos^{-1}(-E_{qm}(\theta ))/\pi -1
\quad .
\end{equation}
That  of course does not imply
that one can manipulate the outcomes of the individual measurements
in order to evoke ``true'' quantum events classically.
I.e., it is impossible to weight or {\em pre}select the individual
events such that the classical expectation function (\ref{e-1}) is
quantum-like.
One encounters parameter dependence but outcome independence
\cite{shimony}.

For example,
suppose again that the two observers measure the sign of the angular
momentum of
$N$ pairs of  classical particles in a singlet state.
But rather than defining their the expectation value by
Eq. (\ref{e-1}), they are {\em weighting} the $N$ normalized individual
outcomes $R^{i}=r_\alpha ^{i}r_\beta ^{i}/N$, $1\le i\le N$, by the
function
\begin{equation}
E_{qqm}(\theta )=-\cos ({\pi \over 2}(R^1+R^2+\cdots +R^N+1))
\quad .
\end{equation}
In this way, a quasi-quantum-like expectation value is defined from
classical events by a global, parameter $\theta$ dependent,
redefinition.

We shall now turn our attention to ---merely hypothetical---
``extremely nonclassical correlations''
and assume a
stronger-than-{\em quantum} expectation function of the form
\begin{equation}
E_s(\theta
)={\rm sgn}
(2\theta /\pi -1)\quad .
\label{e:7}
\end{equation}
$E_s$
can be rewritten in terms of the {\em classical}
expectation function,
\begin{equation}
E_s(\theta
)={\rm sgn}
(E(\theta ))\quad .
\label{e:9}
\end{equation}
$E_s(\theta )$, along with $E(\theta )$ and $E_{qm}(\theta )$,
is drawn in Fig. \ref{f:99}.
 \begin{figure}
 \begin{center}
Figure 2
 \end{center}
\caption{$E_s(\theta )$, $E(\theta )$ and $E_{qm}(\theta )$.
 \label{f:99}}
\end{figure}
Note that, since for $-\pi <x<\pi $,
\begin{eqnarray}
{\rm sgn} (x)
&=&\left\{ \begin{array}{rl}
-1 \; &{\rm  for}\; x<0 \\
 0 \; &{\rm  for}\; x=0 \\
+1 \; &{\rm  for}\; 0< x
 \end{array} \right.\\
&=&{4\over \pi }\sum_{n=0}^\infty {\sin [
(2n+1)x]\over
(2n+1)}
\label{e:8}
\\
&=&{4\over \pi }\sum_{n=0}^\infty (-1)^n{\cos [
(2n+1)(x-\pi /2)]\over
(2n+1)}\quad ,
\label{e:10}
\end{eqnarray}
the quantum mechanical expectation value can be
attributed to
the first summation term in Eq.
(\ref{e:10}).
As a result,
for angles $\pi /2< \theta <\pi$,
the pair of outcomes $+\,+$ or $-\,-$ occurs always,
the pair of outcomes $+\,-$ or $+\,-$ occurs never;
for angles $0< \theta <\pi /2$,
the pair of outcomes $+\,+$ or $-\,-$ occurs never,
the pair of outcomes $+\,-$ or $+\,-$ occurs always.
Under the assumption of 100\% detector efficiency, this cannot be
accommodated
by any classical theory under the assumption of local realism,
nor can we think of any quantum correlation satisfying it.

$E_s(\theta )$ would give rise to a maximal violation of
Bell's
inequality, since in twodimensional polar coordinates $(\varphi , r)$
and for the four directions
$\alpha =(\pi /2,1)$,
$\alpha '=(0,1)$,
$\beta =(\pi /4,1)$,
$\beta '=(3\pi /4,1)$ drawn in Fig. \ref{f-2},
$\vert E(\alpha ', \beta )+E(\alpha , \beta )+E(\alpha ,\beta ')-
E(\alpha ',\beta ')\vert =4$.
 \begin{figure}
 \begin{center}
Figure 3
 \end{center}
\caption{Angle settings for maximal violation of the Bell inequalities.
 \label{f-2}}
\end{figure}
A similar violation of Bell's inequality by the maximal value of $4$
has been studied by
Popescu and Rohrlich
\cite{pop-rohr}
and, for a classical system by Aerts  \cite{aerts}.
It is known \cite{cirelson} that
the maximal violation of Bell's inequality by quantum mechanics
is $2\sqrt{2}$.

Now consider again the
four sheets of paper listing the results of a
series of $N$ experiments.  Fig. \ref{f-8}
demonstrates that one cannot give any listing of outcomes that
would correspond to $E_s(\theta )$. The reason is that
$u(\alpha ',\beta )=
u(\alpha ,\beta )=
u(\alpha ,\beta ')=0$ and
$u(\alpha ',\beta ')=N$
(cf. Fig.
\ref{f-8}).
This implies that
at the same time the lists $\alpha '$ and $\beta '$
are identical, corresponding to the outer path with
$u(\alpha ',\beta ')=N$,
as well as sign reversed, corresponding to the inner path with
$u(\alpha ',\beta )=
u(\alpha ,\beta )=
u(\alpha ,\beta ')=0$.
 \begin{figure}
 \begin{center}
Figure 4
 \end{center}
\caption{Inconsistencies encountered in experiments exhibiting
maximal violation (4) of Bell's inequality.
Again, ``$+$'' stands for $+1$,
``$-$'' stands for $-1$.
 \label{f-8}}
\end{figure}
This results in a complete contradiction.
Thus out of all $N$ particle pairs there is {\em not a single} one
to which a consistent
quadruple of outcomes
can be assigned, which shows that this is a
two-particle analogue to the GHZ setup \cite{ghz}.
In Fig. \ref{f-8} we have illustrated this fact by assigning both
values ($+$ and $-$) to {\em every single} result in list $\beta '$.

The stronger-than-quantum expectation function saturates the Roy-Singh
inequalities \cite{roy-singh} and cannot give rise
to faster-than-light signalling as long as one assumes
unpredictability and/or randomness of the single outcomes
(cf. \cite{pop-rohr}).
Notice that although the two particles are perfectly correlated, the
outcome of the single measurement on either side cannot be controlled
and occurs at random.
An experimenter recording the outcomes for particle one of subsequent
particle
pairs would for instance measure a random sequence
$s_1=++-+-- \cdots$,
 whereas, depending on the relative angle $\theta$, a second
observer, recording the outcomes for the second particle of the
respective
pairs, would measure
either the sequence
$s_2=++-+-- \cdots$
(for $\theta > \pi/2$),
or the sequence
$s_2=--+-++ \cdots$
(for $\theta < \pi/2$).
Since for both experimenters the sequences of outcomes
appear totally uncontrollable and at random
it is impossible to infer the value of $\theta$ on the basis of one of
those sequences alone. This expresses the impossibility of
faster-than-light communication due to the
outcome independence even in case of perfect parameter dependence.

For the sake of completeness, we mention that
{\em weaker-than-classical} expectation functions
\begin{equation}
E_w(\theta )<E(\theta )
\end{equation}
 may be obtained by
adding noise to the classical signal. In its extreme form, the signal is
``washed out'' by noise entirely, such that $E(\theta )=0$. That means
that for all relative angles $\theta $ the pairs of outcomes
$+\,+$,
$-\,-$,
$+\,-$ and
$-\,+$ occur with equal probability.

\subsubsection*{Acknowledgements}
One author (K.S.) acknowledges the patient help of Professor Rainer Dirl
with the quantum mechanics of spin; both authors acknowledge many
discussions with Professor Anton Zeilinger.
This paper has been partly supported by the Austrian Fonds zur
F\"orderung der Wissenschaftlichen Forschung, project nr. P8781-PHY,
Schwerpunkt Quantenoptik.

\subsubsection*{Appendix: Quantum expectation value of two particles of
spin $j$ in a singlet state}
\begin{eqnarray}
C(\theta )&=&
\langle J= 0 ,M= 0\mid \alpha \cdot \hat{J}^A \otimes \beta \cdot
\hat{J}^B\mid
J=0,M=0\rangle
\nonumber
 \\
&=&
 \sum_{m,m'}
\langle  00 \mid jm,j-m \rangle
\langle  jm',j-m'\mid 00 \rangle \times \nonumber
\\
&&
\qquad
\qquad
\qquad
\times
^A\langle jm\mid  ^B\langle j-m\mid
\alpha \cdot \hat{J}^A \otimes \beta \cdot \hat{J}^B
\mid jm'\rangle ^A \mid j-m'\rangle ^B  \nonumber
\\
&=&
 \sum_{m,m'}
\langle  00 \mid jm,j-m \rangle
\langle  jm',j-m'\mid 00 \rangle       \times \nonumber  \\
&&
\qquad
\qquad
\qquad   \times
\langle jm\mid
\alpha \cdot \hat{J}^{A}
\mid jm'\rangle
 \langle j-m\mid
 \beta \cdot \hat{J}^B
 \mid j-m'\rangle                             \nonumber
\\
&=&
 \sum_{m,m'}
{(-1)^{j-m}(-1)^{j-m'}\over 2j+1}
\langle jm\mid
 \hat{J}^A_z
\mid jm'\rangle
\langle j-m\mid
 \beta \cdot \hat{J}^B
 \mid j-m'\rangle \nonumber
\\
&=&
 \sum_{m,m'}
{(-1)^{j-m}(-1)^{j-m'}\over 2j+1}
m \delta _{m m'}
\langle j-m\mid
 \beta \cdot \hat{J}^B
 \mid j-m'\rangle \nonumber
\\
&=&  \sum_m
m{(-1)^{2j-2m}\over 2j+1}
 \langle j-m\mid
 \beta \cdot \hat{J}^B
 \mid j-m\rangle  \nonumber
\\
&=& {1\over 2j+1}  \sum_m
-m^2 \beta_z      \nonumber
\\
&=& -{1 \over 2j+1}\cos \theta  \sum_{m=-j}^j
m^2
\qquad {\rm for} \; 0\le \theta \le \pi
 \nonumber
\\
&=&- {j(j+1)\over 3} \, \cos \theta
\qquad {\rm for} \; 0\le \theta \le \pi
\quad .
 \nonumber
\end{eqnarray}


\clearpage

\begin{thebibliography}{99}

\bibitem{krenn}
G. Krenn, {\sl The probabilistic origin of Bell's inequality}, in
{\sl Proceedings of the 3rd International Workshop on Squeezed States
and Uncertainty Relations}
ed. by
Y.-H. Shih, {\it in print}.

 \bibitem{clauser}
 J. F. Clauser and A. Shimony, {\sl Rep. Prog. Phys.} {\bf 41}, 1881
 (1978).



\bibitem{peres-book}
 A. Peres, {\sl Quantum Theory: Concepts and Methods} (Kluwer Academic
Publishers, Dordrecht, 1993).

\bibitem{peres}
 A. Peres, {\sl Am. J. Phys.} {\bf 46}, 745 (1978).



\bibitem{gisin-peres}
N. Gisin and A. Peres, {\sl Phys. Lett. } {\bf A162}, 15 (1992).


\bibitem{shimony}
 A. Shimony, {\sl Controllable and uncontrollable non-locality}, in
 {\sl Proc. Int. Symp. Foundations of Quantum Mechanics}, ed. by S.
 Kamefuchi {\it et al.} (Physical Society of Japan, Tokyo, 1984);
 A. Shimony, {\sl Events and Processes in the Quantum World}, in {\sl
 Quantum Concepts in Space and Time}, ed. by R. Penrose and C. I. Isham
 (Clarendon Press, Oxford, 1986).

\bibitem{pop-rohr}
S. Popescu and D. Rohrlich,
{\sl Foundations of Physics}
{\bf 24}, 379 (1994).

\bibitem{aerts}
D. Aerts, {\sl Lettere al Nuovo Cimento} {\bf 34}, 107 (1982).
The suggested analogy to two entangled
spin--${1\over 2}$ particles is challenged by the fact that
the proposed expectation functions are {\em not}
invariant with respect to temporal order.
In particular,
 $X_{\alpha \beta}=X_\alpha X_\beta$ if $X_\beta$ is
measured {\em before} $X_\alpha$ but
 $X_{\alpha \beta}\neq X_\alpha X_\beta$ if $X_\beta$ is
measured {\em after} $X_\alpha$.
For incompressible fluids, this would give rise to faster-than-light
signalling and time paradoxa.


\bibitem{cirelson}
B. S. Cirel'son, {\sl Letters in Mathematical Physics}  {\bf 4}, 93
(1980).


\bibitem{ghz}
 D. M. Greenberger, M. Horne and A. Zeilinger, in {\sl Bell's Theorem,
 Quantum Theory, and Conceptions of the Universe}, ed. by M. Kafatos
 (Kluwer, Dordrecht, 1989);
 D. M. Greenberger, M. A. Horne, A. Shimony and A. Zeilinger,
 {\sl Am J. Phys.}
{\bf 58}, 1131 (1990).

\bibitem{roy-singh}
S. M. Roy and V. Singh,
{\sl Phys. Lett.} {\bf A139}, 437 (1989).


\end{thebibliography}
\end{document}